\documentclass[prl,superscriptaddress, preprintnumbers,twocolumn,showpacs]{revtex4}
\usepackage{amsmath}
\usepackage{amssymb}
\usepackage{amsthm}
\usepackage{amsfonts}
\usepackage{algorithmic}
\usepackage{enumerate}
\usepackage{latexsym}
\usepackage{graphicx}
\usepackage{bm}
\usepackage{subfigure}
\usepackage[dvips]{color}

\begin{document}

\title{Localization Trajectory and Chern-Simons Axion Coupling for Bilayer Quantum Anomalous Hall Systems}
\author{Si-Si Wang}
\affiliation{SKLSM, Institute of Semiconductors, Chinese Academy of Sciences, P.O. Box 912, Beijing 100083, China}
\affiliation{College of Materials Science and Opto-Electronic Technology, University of Chinese Academy of Sciences, Beijing 100049, China}
\affiliation{Synergetic Innovation Center of Quantum Information and Quantum Physics, University of Science and Technology of China, Hefei, Anhui 230026, China}
\author{Yan-Yang Zhang}
\email{yanyang@gzhu.edu.cn}
\affiliation{School of Physics and Electronic Engineering, Guangzhou University, 510006
Guangzhou, China}
\author{Ji-Huan Guan}
\affiliation{SKLSM, Institute of Semiconductors, Chinese Academy of Sciences, P.O. Box 912, Beijing 100083, China}
\affiliation{School of Physical Sciences, University of Chinese Academy of Sciences, Beijing 100049, China}
\affiliation{Synergetic Innovation Center of Quantum Information and Quantum Physics, University of Science and Technology of China, Hefei, Anhui 230026, China}
\author{Yan Yu}
\affiliation{SKLSM, Institute of Semiconductors, Chinese Academy of Sciences, P.O. Box 912, Beijing 100083, China}
\affiliation{School of Physical Sciences, University of Chinese Academy of Sciences, Beijing 100049, China}
\affiliation{Synergetic Innovation Center of Quantum Information and Quantum Physics, University of Science and Technology of China, Hefei, Anhui 230026, China}
\author{Yang Xia}
\affiliation{Microelectronic Instrument and Equipment Research Center, Institute of Microelectronics of Chinese Academy of Sciences, Beijing 100029, China}
\affiliation{School of Microelectronics, University of Chinese Academy of Sciences, Beijing 100049,
China}
\author{Shu-Shen Li}
\affiliation{SKLSM, Institute of Semiconductors, Chinese Academy of Sciences, P.O. Box 912, Beijing 100083, China}
\affiliation{College of Materials Science and Opto-Electronic Technology, University of Chinese Academy of Sciences, Beijing 100049, China}
\affiliation{Synergetic Innovation Center of Quantum Information and Quantum Physics, University of Science and Technology of China, Hefei, Anhui 230026, China}
\date{\today}

\begin{abstract}
Quantum anomalous Hall (QAH) multilayers provide a platform for topological materials with high Chern numbers. We investigate the localization routes of bilayer QAH systems with Chern number $C=2$ during the process of increasing disorder, by numerical simulations on their quantum transport properties and the Chern-Simons axion coupling. The localization trajectories present richer behaviors than those in the monolayer with $C=2$. For example, there exists a stable intermediate state with $C=1$ before localization, which was always unstable in a monolayer. In some cases, this $C=1$ state is ``weak'' in the sense that its Hall plateau is hardly visible in a mesoscopic sample, but is still stable in the sense of renormalization group. The underlying physics is discussed. During the localization process, the Chern-Simons axion coupling shows a surprising peak which is even more remarkable in the large size limit. The physical origin of this peak is understood by a real space analysis of the electronic states. As a result, the disordered QAH multilayers can be good candidates for this nontrivial magnetoelectric coupling mediated by orbital motions.
\end{abstract}

\maketitle

\section{I. Introduction}
The research of topological materials recovers the interests of searching the quantum anomalous Hall (QAH) effects, i.e.,
Chern insulators without an external magnetic field\cite{Haldane}.
QAH effects in different materials have been theoretically proposed \cite{QAH_CXLiu,QAH_HJZhang,QAH_QZWang,QAH_RYu,QAH_SCWu} and
experimentally observed \cite{QAH_Exp1,QAH_Exp2,QAH_Exp3,QAH_Exp4,QAH_Exp5}.
The Chern insulator, as a topological state, is robust against weak disorder, manifested as a plateau of Hall conductance $\sigma_{xy}=\frac{e^2}{h}C$ with $C$ the Chern number, and a vanishing longitudinal conductance $\sigma_{xx}$\cite{Klitzing1980,Thouless1982,QNiuRMP}. Nevertheless, it can still be localized into Anderson insulators with $C=0$ ultimately, by
sufficiently strong disorder. This can be regarded as a disorder induced topological transition with Chern number from nonzero to zero\cite{QHEDisorder,YYZhang2012,ZG2015}. On the other hand, this trajectory towards localization can also be understood in the context of scaling and renormalization group (RG), i.e., the Pruisken formalism\cite{Pruisken1985,Huckestein1995}, where all the quantized Hall conductances (associated with all nonzero Chern numbers) correspond to stable fixed points attracting all nearby RG flows in the conductance plane $(\sigma_{xy},\sigma_{xx})$. Besides the analytical RG treatment, these trajectories along scaling transformation can also be observed by experiments\cite{RG_Exp2,Checkelsky2014,SGrauer2017} or simulated by numerical calculations\cite{Xue2013,Werner2015,ZG2015,YXXing2018}.

QAH states with different Chern numbers are topologically distinct from each other. The localization trajectory of the monolayer (two-band) QAH state with Chern number $C=2$ has been numerically investigated\cite{Song2015,ZG2015}.
The novelty is that the intermediate state with $C=1$ is not stable during the process towards localization, no matter how the model parameters are tuned. Microscopically this has been attributed to a $C=1$ state rooted in broad statistical tails, which cannot manifests itself after statistical averaging and size scaling\cite{ZG2015}.

In additional to transport properties, the orbital mediated magnetoelectric coupling attracts many attentions in topological materials. The axion electrodynamics\cite{Wilczek1987} introduces a non-Abelian Lagrangian term\cite{Spaldin2005,Fiebig2005,Hehl2008,Essin2009,Olsen2017},
\begin{equation}
\mathcal{L}_{\theta}={\frac{e^2}{h}}{\frac{\theta}{2\pi}} {\bm{E}\cdot\bm{B}},\label{EqAxion}
\end{equation}
where $\theta$ is a dimensionless parameter which is determined by the electronic structure of the material, and $\bm{E}$ and $\bm{B}$ are the electric and magnetic fields respectively. This quantity reflects a deep magnetoelectric connection mediated by the electronic orbital motions in solids, especially remarkable in topological materials.
Due to the involvement of three dimensional motions [see Eq. (\ref{EqTheta0}) below], a bilayer system, which will be investigated in this paper, is the simplest lattice with a nonzero Chern-Simons axion coupling\cite{Olsen2017}.

An important knowledge from (quasi) two-dimensional (2D) systems is that
a bilayer system may possess quite different properties from its monolayer counterpart\cite{GrapheneRMP,QHE_Bilayer}. Few layer QAH systems have been experimentally fabricated recently by molecular beam epitaxy\cite{QAH_Multilayer}. The bilayer QAH system with $C=2$ we will discuss is essentially different from a monolayer one in at least two aspects. First, it is a four band model, or, a model with two pairs of bands, with each pair contributing one Chern number. Second, each pair corresponds to one layer which is distinguishable in the real space, and their coupling is tunable.

In this paper, we will investigate the trajectory towards localization of such a bilayer system with increasing disorder, by performing numerical simulations on the two terminal conductance, two-parameter scaling, and the axion magnetoelectric coupling. We find that, there is a stable Hall conductance plateau with $C=1$, in a region of intermediate disorder strength. For some model parameters, this state may be hardly visible (imperfect plateau of $\sigma_{xy}$ and nonzero $\sigma_{xx}$ ) at finite sample size, but scaling shows that it is still on the way towards $(\sigma_{xy}=1\times\frac{e^2}{h},\sigma_{xx}=0)$. The existence of such a stable intermediate Chern state has been ruled out for single layer QAH systems\cite{Song2015,ZG2015}. Through numerical scaling, we also find that some trajectories cannot be perfectly fitted into the Pruisken flows, especially in the limit of strong inter-layer coupling. Finally, the disorder induced collapse of the quantum Hall plateau is shown to correspond to a peak of axion magnetoelectric coupling.

\section{II. The Model}

\begin{figure}[htbp]
\includegraphics*[width=0.4\textwidth]{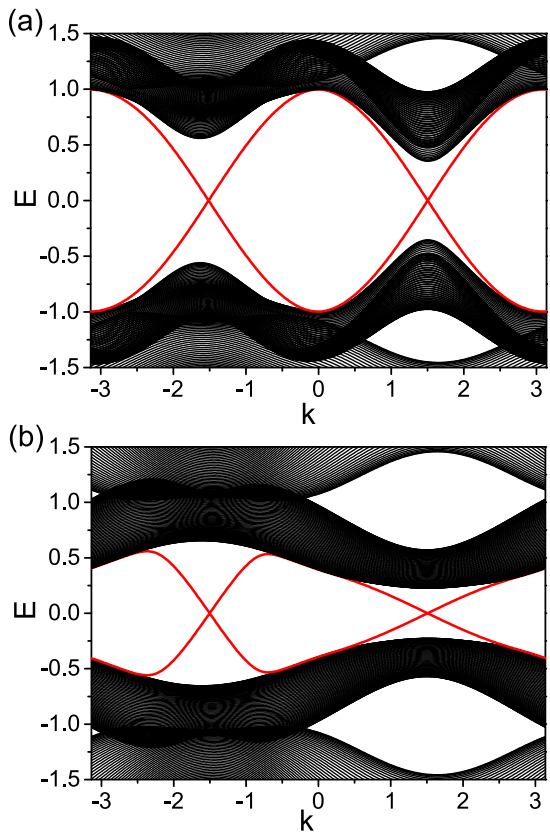}
\caption{(Color online) The band structure of the quasi-one
dimensional ribbon with width $N_y=80$ in the clean limit, for (a) $A_1=1$ and (b) $A_1=0.3$.
The rest model parameters are identical for both panels: $A_1=0.3, B_1=0.2, D_1=0,
M_1=0.2, A_2=1.0, B_2=0.6, D_2=0, M_2=1.0$, and $t=0.3$. The red lines are edge states.}
\label{FigDispersion}
\end{figure}

The general form of a spinless bilayer QAH system can be expressed as
\begin{equation}
\mathcal{H}_{\mathrm{bi}}=\mathcal{H}_1+\mathcal{H}_2+\mathcal{H}_c,\label{EqTotalHamiltonian}
\end{equation}
where $\mathcal{H}_L$ ($L=1,2$) is the Hamiltonian for the $L$-th layer, and
$\mathcal{H}_c$ is the coupling between them. Here, we choose each layer to be the spin up
component of the Bernevig-Hughes-Zhang (BHZ) model\cite{Bernevig2006} defined on a square lattice, with one $s$ orbital and one $p$ orbital on each site. This is one of the minimum models of the QAH state. In the $k$ space, the layer Hamiltonian reads
\begin{equation}
\mathcal{H}_L=\sum_{\bm{k,\alpha\beta}}c^{\dagger}_{L;\bm{k}\alpha}H_{L;\alpha\beta}(\bm{k}) c_{L;\bm{k}\beta},\label{EqHL0}
\end{equation}
where
$c^{\dagger}_{L;\bm{k}\alpha}$ ($c_{L;\bm{k}\alpha}$) creates (annihilates) an electron with wavenumber $\bm{k}$ and orbital $\alpha\in \{s,p\}$ in layer $L\in\{1,2\}$. For BHZ model, $H_{L;\alpha\beta}(\bm{k})$ is a $2 \times 2$ matrix as\cite{Bernevig2006}
\begin{eqnarray}
H_L(\bm{k})&=&\varepsilon_L(\bm{k})I_{2\times 2}+\sum_{i}d^i_L(\bm{k})\sigma_i \label{EqBHZ}\\
\varepsilon_L(\bm{k})&=&-2D_L\big[2-\cos ( k_{x}-q_L^x )-\cos ( k_{y}-q_L^y )\big] \nonumber \\
d^1_L(\bm{k})&=&A_L\sin ( k_{x}-q_L^x ), \quad d^2_L(\bm{k})=A_L\sin ( k_{y}-q_L^y )\nonumber\\
d^3_L(\bm{k})&=&M_L-2B_L\big[2-\cos (k_{x}-q_L^x) -\cos ( k_{y}-q_L^y)\big],\nonumber
\end{eqnarray}
where $\sigma_i$ are the Pauli matrices acting on the orbital space $\{s,p\}$.
We have included a layer dependent momentum shift $\bm{q}_L=(q_L^x,q_L^y)$ of the band structure in the Brillouin zone, so that two groups of edge states (originated from two layers) can be better distinguished visually, as can be seen in Fig. \ref{FigDispersion} (a). We have checked that this shift does not affect the transport properties investigated in the following.
The real space version of the layer Hamiltonian $\mathcal{H}_L=\sum_{ij,\alpha\beta}c^{\dagger}_{L;i\alpha}H_{L;\alpha\beta}(i,j) c_{L;j\beta}$ can be obtained from Eqs. (\ref{EqHL0}) and (\ref{EqBHZ})
by performing a straightforward inverse Fourier transformation
$c_{L;\bm{k}\beta }=\frac{1}{\sqrt{V}}\sum_{j}c_{L;j\beta }e^{-i\bm{k}\cdot {
\bm{x}_{j}}}$, where $i$ is the site index. The size of such a finite sample is characterized by the length $N_x$ and the width $N_y$ measured in units of the lattice constant.

In the absence of the inter-layer coupling $\mathcal{H}_c$, the band structure and Chern number of
each layer can be tuned independently by varying the parameters.
For example, the band gap is $2|M_L|$, and the layer Chern number
\begin{equation}
C_L=
\left\{
   \begin{array}{lll}
      +1,&\quad 0<M_L/2B_L<2 \\
      0,&\quad M_L/2B_L<0\\
      -1,&\quad \mathrm{otherwise}. % To SS: insert the condition here.
    \end{array}
\right.
\end{equation}
The inter-layer coupling is considered to be the simple
form in real space as
\begin{equation}
\mathcal{H}_c=\sum_{i\alpha}\big(t c^{\dagger}_{1;i\alpha}c_{2;i\alpha}+\mathrm{H. c.}\big),
\label{EqHc}
\end{equation}
where $t$ is the strength of the coupling.
Without inter-layer coupling, the Chern number of this ``bilayer''
system is trivially $C=C_1+C_2$.
Consequently, if the adiabatic turning on of the inter-layer coupling term (\ref{EqHc}) does not close the bulk gap, the Chern number $C$ of the bilayer system will not change. Throughout this paper, we focus on this case with $C=2$, constructed from coupling two layers with $C_L=1$.

In Fig. \ref{FigDispersion}, we present the band structures of the bilayer systems in the ribbon geometry, with two typical groups of model parameters. Two groups of edge states (red lines) in the bulk gap can be clearly seen, reflecting the Chern number $C=2$ at half filling. To have a direct visual comparison with the case of a single layer\cite{Song2015,ZG2015}, we have appropriately tuned the model parameter $\bm{q}_L$ [see Eq. (\ref{EqBHZ})], so that two groups of edge states (originated from two layers respectively) are centered at different locations with two local gaps respectively. Furthermore, the magnitudes of these two local gaps can be similar [Fig. \ref{FigDispersion} (a), with $A_1=1.0$] or different [Fig. \ref{FigDispersion} (b) $A_1=0.3$]. In the following, we will see that these two patterns correspond to different quantum transport properties.

The effect of disorder is modeled in real space by adding a random onsite potential term
\begin{equation}
\sum_{L;i\alpha} c^{\dagger}_{L;i\alpha} U_{L;i} c_{L;i\alpha}\label{EqRandomOnsite}
\end{equation}
to the Hamiltonian $\mathcal{H}_{\mathrm{bi}}$, where $U_{L;i}$ are independent random numbers uniformly distributed in $(-\frac{W}{2},\frac{W}{2})$, and $W$ is the disorder strength.

\section{III. Methods}

At zero temperature, the two-terminal conductance of a finite sample
can be expressed by Green's functions as\cite{Datta,Khomyakov2005,Caroli2005}
\begin{equation}
G=\frac{e^{2}}{h}\mathrm{Tr}\left[  \Gamma_{S}G^{r}\Gamma
_{D}G^{a}\right]  , \label{EqConductance}%
\end{equation}
where
$G^{r/a}$ is the retarded/advanced Green's function, and $\Gamma_{S(D)}%
=i(\Sigma_{S(D)}^{r}-\Sigma_{S(D)}^{a})$ with $\Sigma_{S(D)}%
^{r/a}$ being retarded/advanced self energies due to the source (drain) lead, respectively.
For a Chern insulator, this two-terminal conductance just equals
the Hall conductance, $G=\sigma_{xy}$\cite{Datta}.
To plot the two-parameter scaling trajectories, longitudinal and transverse conductances should be calculated respectively, which will be described below.

For a finite 2D sample with disorder, the concept of the wavenumber $\bm{k}=(k_x,k_y)$ can be restored if twisted boundary conditions are adopted along both directions\cite{QNiu1984,QNiuRMP,QHEDisorder,YYZhang2012,ZG2015}. This is simply equivalent to treating this finite sample as the supercell of an infinite superlattice. In this sense, hereafter, we call these finite samples as supercells and the energy bands of this superlattice as subbands. Now, the longitudinal conductance $\sigma_{xx}$ of the supercell is calculated
as the Thouless conductance\cite{Thouless1972,Braun1997}
\begin{equation}
\sigma_{xx}(E_n)=\frac{e^{2}}{h}\pi\rho(E_n)\frac{\partial^2 E_n}{\partial k_x^2}\Big|_{k_x=0}, \label{EqSigmaxx}
\end{equation}
where $\rho$ is the density of states, and $E_n$ is the subband at the Fermi energy.

In this context of disordered supercell, the (intrinsic) Hall conductance $\sigma_{xy}$
is found to be proportional to the integration of the Berry curvature $F^z_{n}$ over the Brillouin zone (BZ) under the Fermi energy $E_F$\cite{Simons1983,Kohmoto1985,QNiuRMP,Werner2015,AHERMP}
\begin{eqnarray}
\sigma_{xy}(E_n)&=\frac{e^{2}}{h}&\sum_{n}{\frac{1}{2\pi i}}\int_{BZ} d^2\bm{k} F^z_{n}\big(\bm{k}\big)f_0\big[E_n(\bm{k})\big] \label{EqSigmaxy}\\
\bm{F}_n&=&\nabla_{\bm{k}} \times \bm{A}_n \nonumber\\
\bm{A}_n&=&\big\langle n(\bm{k})|\nabla_{\bm{k}}|n(\bm{k})\big\rangle,\nonumber
\end{eqnarray}
where $f_0(E)$ is the Fermi distribution at zero temperature, and $\big|n(\bm{k})\big\rangle$ is the normalized wave function of the $n$-th subband which satisfies $H(\bm{k})\big|n(\bm{k})\big\rangle=E_n(k)|n(\bm{k})\rangle$.
If $E_F$ is within the bulk gap,
$\sigma_{xy}=\frac{e^2}{h}C$ with $C$ the total Chern number of occupied (sub)bands, where $\sigma_{xy}$ carried by the edge states and $C$ determined by the bulk bands are related through the bulk-boundary correspondence\cite{Thouless1982,QNiuRMP}. The Hall conductance $\sigma_{xy}$ may not be quantized if some subbands are only partly filled\cite{QNiuRMP,AHERMP}. This relation between Berry curvature and Hall conductance has been used to design and control the anomalous Hall effect in magnetic Heusler
compounds\cite{KManna2018}. Our numerical evaluations of Eqs. (\ref{EqSigmaxx}) and (\ref{EqSigmaxy}) are based on the methods used in previous works\cite{Fukui2004,Werner2015,ZG2015}.
Hereafter, all conductances will be expressed in units of $\frac{e^2}{h}$.

In the presence of disorder, an ensemble average of these transport quantities over disorder configurations should be performed to obtain physically meaningful results, for a certain group of model parameters and supercell size. Then, by enlarging the supercell size, the ensemble averaged conductance vector $\bm{\sigma}\equiv (\sigma_{xy},\sigma_{xx})$ forms a flow in this 2D parameter space\cite{Werner2015,Huckestein1995,ZG2015}.
Throughout this paper, we will call this a ``scaling flow'' (or a ``scaling trajectory''), while that calculated from RG theories\cite{Pruisken1985} an ``RG flow'' (or an ``RG trajectory''). These flows offer the information of transport properties in the thermodynamic limit.

Now we introduce the calculation of orbital magnetoelectric coupling. Due to the close relation between $\theta$ in Eq. (\ref{EqAxion}) and the linear orbital magnetoelectric response $\alpha_{ij}=(\partial P_i/\partial B_j)_{\bm{E}}=(\partial M_j/\partial E_i)_{\bm{B}}$, with $\bm{P}$ and $\bm{M}$ the macroscopic polarization and magnetization\cite{Olsen2017}, it is
adequate to focus on the Chern-Simons axion (CSA) coupling\cite{Essin2009,Olsen2017}
\begin{equation}
\bar{\alpha}_{\mathrm{CS}}={\frac{e^2}{h}}{\frac{\theta_{\mathrm{CS}}}{2\pi}},
\end{equation}
where $\theta_{\mathrm{CS}}$ is a part of $\theta$ with a $2\pi$ ambiguity.
For an infinite three-dimensional (3D) crystalline material,
the CSA coupling can be expressed as\cite{XLQi2008}
\begin{equation}
\theta_{\mathrm{CS}}=-\frac{1}{4\pi}\int d\bm{k}\epsilon_{ijl} \mathrm{Tr}\big(A^i_{\bm{k}}\partial_{k_j}A^l_{\bm{k}}-i\frac{2}{3}A^i_{\bm{k}}A^j_{\bm{k}}A^l_{\bm{k}}\big),
\label{EqThetaCS}
\end{equation}
where $A^i_{\bm{k}}$ is the matrix of Berry connection between subbands in direction $i$, and the trace is over occupied bands.

Moreover, the expression of the CSA coupling has been carefully generalized to an infinite 2D slab as $\theta_2$, and to a finite (zero dimensional) crystal as \cite{Sinisa2009,Malash2010,Marzari1997,Olsen2017}
\begin{equation}
\theta_0=-8\pi^2\mathrm{Im}\mathrm{Tr}[P\mathit{x}P\mathit{y}P\mathit{z}]\label{EqTheta0},
\end{equation}
where $P=\sum_n|\psi_n\rangle\langle\psi_n|$ is the projection operator onto the occupied subspace in the ground state.
It has been verified that in the thermodynamic limit\cite{Olsen2017},
\begin{equation}
\frac{\theta_0}{N_x N_y}\rightarrow \theta_2, \quad\mathrm{for}\quad N_x,N_y \rightarrow \infty
\label{EqTheta02}
\end{equation}
This $\theta_0$ defined by Eq. (\ref{EqTheta0}) is convenient for treating disordered and finite samples which we are interested here.
Moreover, $\theta_0$ is demonstrated to be a more fundamental definition of CSA than those in $k$ space\cite{Olsen2017}.
The CSA coupling [Eqs. (\ref{EqThetaCS}) and (\ref{EqTheta0})] involves responses in all three directions and thus it is meaningless for single layer systems.

\begin{figure}[htbp]
\includegraphics*[width=0.35\textwidth]{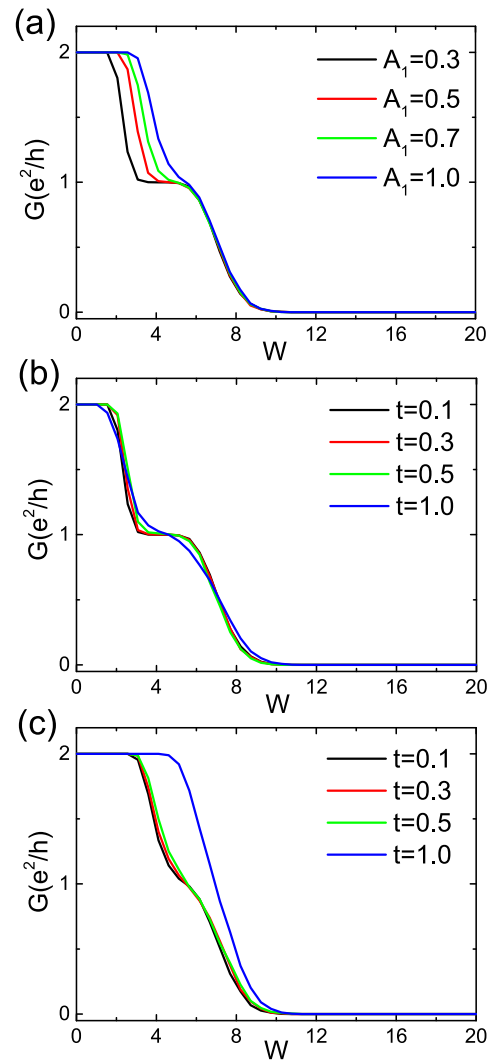}
\caption{(Color online) Disorder averaged two-terminal conductance $G$ as a function of disorder strength $W$, at Fermi energy $E_F=0.05$.
(a) Fixing $t=0.3$ while changing $A_1$. (b) Fixing $A_1=0.3$ while changing $t$. (c) Fixing $A_1=1$ while changing $t$. The blue and black curves in panel (a) correspond to the cases of Fig. \ref{FigDispersion} (a) and (b) respectively. The sample size is $100\times100$ and the average is over 500 disorder configurations. The rest model parameters are identical to Fig. \ref{FigDispersion}. }
\label{FigConduct}
\end{figure}

\section{III. Results }

\subsection{A. Appearance of the intermediate state with $C=1$ }

A Chern insulator (with $C\neq0$) manifests itself as a robust Hall conductance plateau at weak disorder, but it can be localized at strong disorder. For a monolayer QAH system (with two bands) with Chern number $C=2$, it has been found that,
the intermediate state with $C=1$ is not a stable plateau on the route towards localization\cite{Song2015,ZG2015}.
Here we try to test the stability of the intermediate state $C=1$ of the bilayer system (\ref{EqHL0}).
Due to the adjustability of each constituent layer, we can make these two layers (especially their gaps) quite asymmetric to each other, by tuning model parameters distinct on two layers.

To have a first glance at the quantum transport property in the presence of disorder, we calculate the two-terminal conductance by using Eq. (\ref{EqConductance}), which is equal to the Hall conductance for a Chern insulator\cite{Datta}. In Fig. \ref{FigConduct}, the disorder averaged conductance $G$ is plotted as a function of the disorder strength $W$, with different $A_1$ (tuning layer gaps), or different $t$ (tuning the inter-layer coupling). Fig. \ref{FigConduct} (a) is the results with fixed $t=0.3$ but different $A_1$. When $A_1=1.0$ (blue line), corresponding to the case of similar local gaps shown in Fig. \ref{FigDispersion} (a), the conductance decays from 2 to 0 persistently, without a visible intermediate plateau with $G=1$, similar to the monolayer case\cite{ZG2015}. However, with decreasing $A_1$, so that two local gaps are more and more different [see Fig. \ref{FigDispersion} (b)], an intermediate plateau with $G=1$ emerges gradually. This plateau with $G=\sigma_{xy}=1$ seems shorter if the inter-layer coupling $t$ is very large, as shown in In Fig. \ref{FigConduct} (b). On the other hand, in the case of similar local gaps ($A_1=1$), as one can imagine, changing $t$ will never give rise to a stable plateau with $\sigma_{xy}=1$.

\begin{figure*}[htbp]
\includegraphics*[width=0.72\textwidth,bb=95 86 690 540]{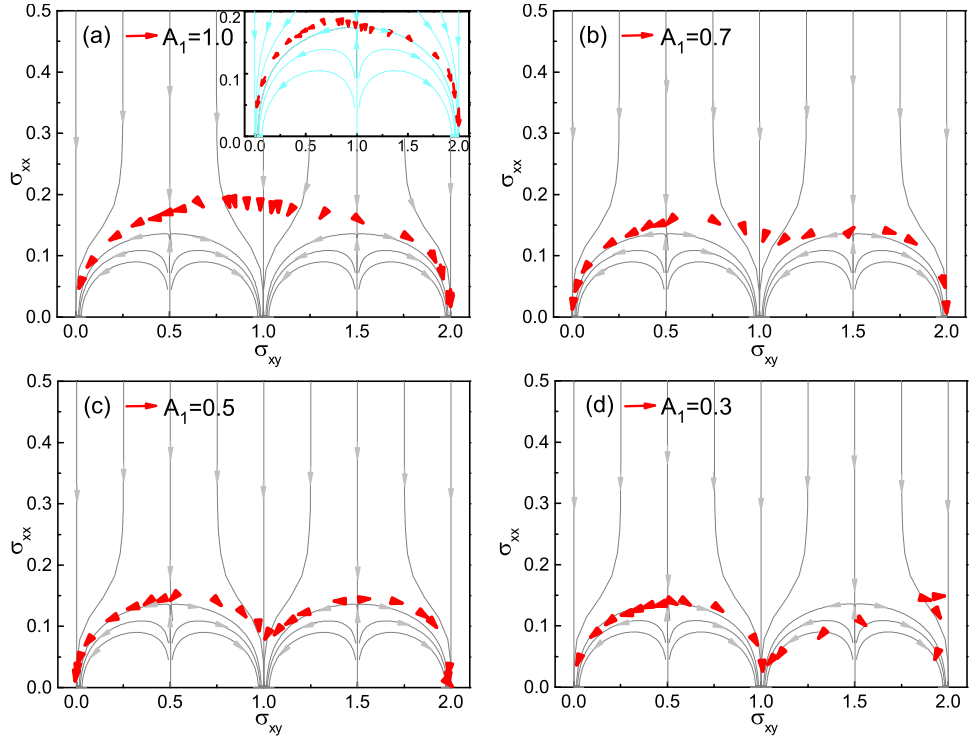}
\caption{(Color online) Red arrows are supercell size driven scaling flows for $t=0.3$, with sample sizes $10\times 10 \rightarrow 15\times 15$: (a) $A_1=1.0$, (b) $A_1=0.7$,
(c) $A_1=0.5$, (d) $A_1=0.3$, which respectively correspond to four curves in Fig. \ref{FigConduct} (a). Each arrow is an average over 10000 to 30000 disorder configurations. The disorder strength $W$ varies from 0 to 12, and the Fermi energy $E_F=0.05$ near the gap center. Grey (main panels, for class $\Gamma_{\mathrm{T}}$) and blue (inset, for class $\Gamma_{\mathrm{R}}$) curves are illustrative configurations of Pruisken RG trajectories\cite{Pruisken1985,Olsen2017} for fitting (see text for details).}
\label{FigRGA1}
\end{figure*}

\begin{figure*}[htbp]
\includegraphics*[width=0.72\textwidth,bb=95 86 690 540]{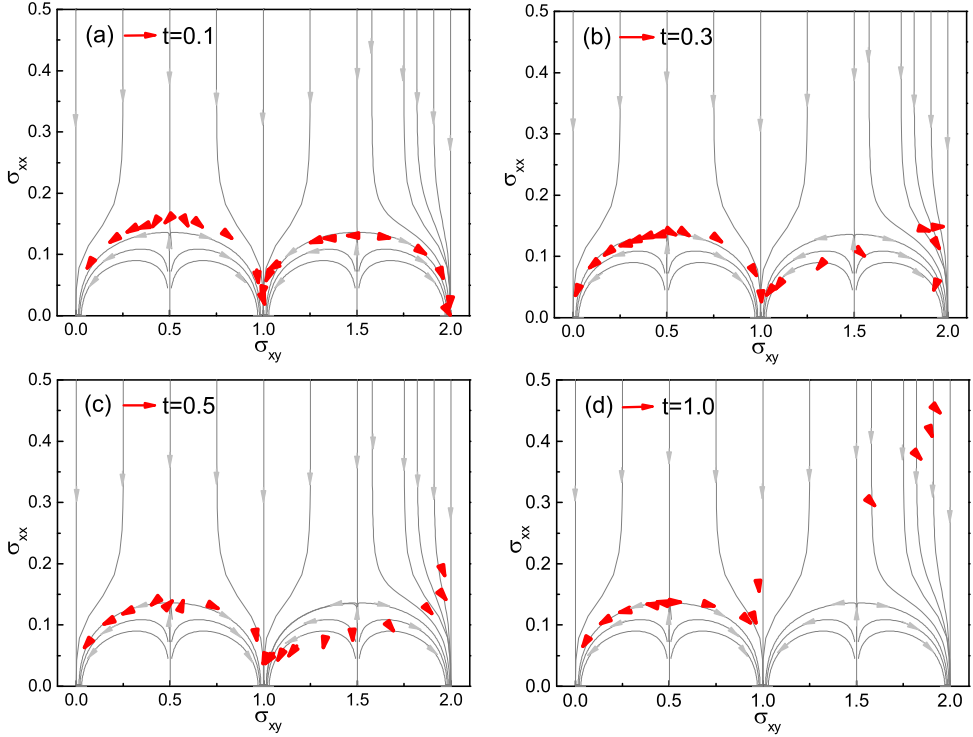}
\caption{(Color online) Similar to Fig. \ref{FigRGA1} but for fixed $A_1=0.3$ and different $t$: (a) $t=0.1$, (b) $t=0.3$,
(c) $t=0.5$, (d) $t=1.0$, which correspond to the cases of four curves shown in Fig. \ref{FigConduct} (b), respectively. }
\label{FigRGt}
\end{figure*}

\subsection{B. Scaling flows }

In Pruisken's pioneering work based on field theory\cite{Pruisken1985}, the RG flows of the integer quantum Hall effect on the $\bm{\sigma}\equiv(\sigma_{xy},\sigma_{xx})$ plane is found to form a periodic pattern, which is illustrated as grey curves in Figs. \ref{FigRGA1} and \ref{FigRGt}.
For these RG flows, $\bm{\sigma}= (n,0)$ are stable fixed points attracting all surrounding flows, corresponding to the robust $n$-th QHE plateau states.
On the other hand, $\bm{\sigma}= (n+\frac{1}{2},\sigma^*_{xx})$ (with $\sigma^*_{xx}\sim 1$) are saddle points (unstable fixed points) practically repelling most surrounding flows. There are main semicircles connecting neighboring stable and saddle points.
It is observed that this nested hierarchical structure of fixed points
can be described by the modular group $\Gamma(1)$, which is infinite,
discrete, and non-Abelian. Indeed, a very wide range of experimental quantum Hall data of scaling flows have been found to perfectly match just a few universality classes, and each class corresponds to a certain subgroup of the modular group\cite{KSOlsen2018}.
In this context, the above Pruisken picture corresponds to the class of $\Gamma_{\mathrm{T}}$\cite{KSOlsen2018}.

Another class of RG flow, $\Gamma_{\mathrm{R}}$, is just a size doubling of the above $\Gamma_{\mathrm{T}}$ flow pattern\cite{KSOlsen2018}, with even Chern number states $\bm{\sigma}= (2n,0)$ as stable fixed points, and with odd Chern number states $\bm{\sigma}= (2n+1,2\sigma^*_{xx})$ as saddle points [See the blue curves in the inset of Fig. \ref{FigRGA1} (a)].
Numerical simulations have shown that the scaling flows of the monolayer QAHE with $C=2$ match this $\Gamma_{\mathrm{R}}$ pattern, with flows going \emph{away} from $\bm{\sigma}= (1,2\sigma^*_{xx})$\cite{ZG2015}.

In order to test the modular symmetry of our bilayer model, we plot the supercell size driven scaling flows on the $\bm{\sigma}$ plane, by numerical simulations of conductances using Eqs. (\ref{EqSigmaxx}) and (\ref{EqSigmaxy}), as in previous works\cite{Werner2015,ZG2015,YXXing2018}. We remind again that in this paper, the flows from the RG techniques and from the conductance simulations are termed as ``RG flows'' and ``scaling flows''
respectively.
The results of scaling flows for different $A_1$ are displayed in Fig. \ref{FigRGA1} as red arrows, where each data point is an average over at least 10000 disorder configurations (30000 configurations around transitions). For comparison, RG flow curves of the Pruisken pattern\cite{Pruisken1985} are also
illustrated in Fig. \ref{FigRGA1}. These RG flows are just plotted by fitting the original RG equations\cite{Pruisken1985} with a few free parameters (including zooming factors in two directions), instead of by a rigorous RG treatment of our bilayer model.
Due to the critical slowing down and large fluctuations near the transitions, it is rather difficult to have a precise and global reproduction (or prediction) of the RG flows from numerical scaling flows\cite{Huckestein1995}. However, intuitive insights can still be drawn.

Fig. \ref{FigRGA1} (a) is the scaling flows for the case of $A_1=1$, with two local gaps similar. Here, we simultaneously fit the same group of scaling flows (red arrows) to the RG flows of $\Gamma_{\mathrm{T}}$ class (grey curves in the main panel) and to those of
$\Gamma_{\mathrm{R}}$ class (blue curves in the inset), respectively.

The profile of this large semicircle of red arrows seems to follow the RG pattern of the modular symmetry $\Gamma_\mathrm{R}$ [blue curves in inset of Fig. \ref{FigRGA1} (a)], i.e., with $\bm{\sigma}= (1,2\sigma_{xx}^*)$ a saddle point \cite{KSOlsen2018}, as in the monolayer case of $C=2$\cite{Song2015}. However, something is different if we scrutinize the orientations of the red arrows around $\sigma_{xy}=1$: Although far away from $\bm{\sigma}= (1,0)$, they show an obvious tendency \emph{towards} $\sigma_{xy}=1$, which is opposite to the $\Gamma_\mathrm{R}$ RG flows (blue curves in the inset) going \emph{away} from $\sigma_{xy}=1$ on the main semicircle. Instead, the orientations of these red scaling arrows around $\sigma_{xy}=1$ comply more with the $\Gamma_\mathrm{T}$ RG flows in the main panel, within the waterfall converging towards $\bm{\sigma}_{1}\equiv (1,0)$. In brief, the scaling flows around $\sigma_{xy}=1$ are more likely to match the RG flows of class $\Gamma_{\mathrm{T}}$ instead of class $\Gamma_{\mathrm{R}}$, although some of the scaling flows are not on the main semicircles.

\begin{figure}[t]
\includegraphics*[width=0.47\textwidth]{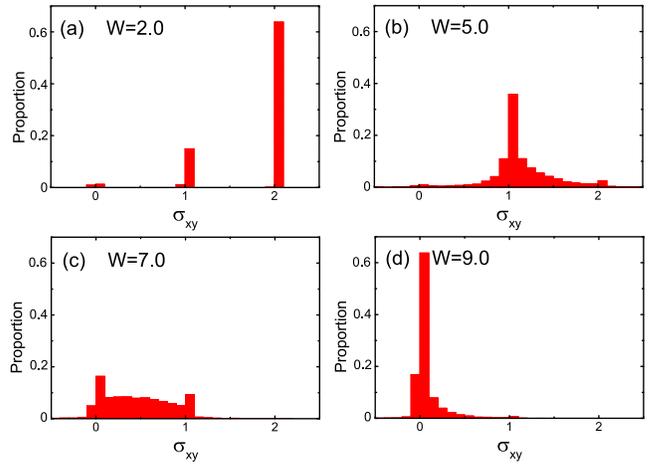}
\caption{Distribution of the Hall conductance $\sigma_{xy}$ corresponding to the case of Fig. \ref{FigRGA1} (a) with $t=0.3$ and $A_1=1$, where the state with $\sigma_{xy}=1$ is weak. Four panels correspond to different disorder strengths $W$: (a) $W=2.0$; (b) $W=5.0$; (c) $W=7.0$; (d) $W=9.0$. }
\label{FigDistribute2}
\end{figure}

Here in Fig. \ref{FigRGA1} (a), we meet an example of badly observable $C=1$ state in the modular symmetry class $\Gamma_{\mathrm{T}}$, manifested as non-quantized $\sigma_{xy}$ and nonzero $\sigma_{xx}$ at finite sample size. This can be attributed to some residual bulk states in the system [see following analysis on Fig. \ref{FigDistribute2} (b)]. Since this state is still on the route towards the stable fixed point $\bm{\sigma}=(1,0)$, a quantized Hall plateau should be observed in an \emph{extremely} large sample, which may significantly exceed the quantum coherence length in practice. We call such Chern states hardly visible in a mesoscopic sample (although it is stable in the RG sense) as a ``weak quantum (anomalous) Hall effect''.

Another issue should be addressed here. In Pruisken's formalism\cite{Pruisken1985}, there is always a definite flow vector at any position of the upper half of the $(\sigma_{xy},\sigma_{xx})$ plane, except at isolated fixed points. For an experiment or a numerical simulation, one cannot predict a priori
where a certain flow will start\cite{Huckestein1995,KSOlsen2018}. In other words, the key information of the scaling flows is the orientation configuration at different positions, instead of the positions themselves, because the Pruisken theory does not prohibit or prefer any position for the flows to appear. Indeed, in many experimental results, the scaling flows are not on the main semicircles\cite{Huckestein1995,Checkelsky2014,KSOlsen2018}.

When $A_1$ is decreased, in Fig. \ref{FigRGA1} (b) to (d) , with two local gaps different, the state with $\sigma_{xy}=1$ becomes more and more attractive for scaling flows, suggesting that it is a stable state during the route towards localization.

The emergence of this stable intermediate plateau with $\sigma_{xy}=1$ is the first finding of this work. Since this state is much more stable and significant when the magnitudes of two local gaps (each accommodating a group of edge states) are extremely different [Fig. \ref{FigDispersion} (b)], this can be illustrated as disorder induced band inversion \cite{YYZhang2012,ZG2015} at the smaller local gap. This is completely different from the monolayer QAH effect with $C=2$\cite{Song2015,ZG2015}. In that case, changing model parameters, e.g., making two local gaps (each of which carries a group of edge states) different will never give rise to a stable plateau with $\sigma_{xy}=1$, and the scaling flows around $\sigma_{xy}=1$ are going away from it. \cite{ZG2015}.

Another notable feature can be seen in the most asymmetric case with small $A_1=0.3$ [Fig. \ref{FigRGA1} (d)]: besides straying away from the main semicircles between $\sigma_{xy}=1$ and $\sigma_{xy}=2$, the orientations of these scaling flows seem to deviate from the typical Pruisken RG configuration\cite{Pruisken1985,KSOlsen2018}. This tendency becomes more remarkable when the inter-layer coupling $t$ is increased, as can be seen in Fig. \ref{FigRGt} (c) and (d).
For example, the orientations of the scaling flows remarkably deviate from those of the standard Pruisken RG flows, especially when $\sigma_{xy}\gtrsim 1.5$. When $\bm{\sigma}$ is high above the semicircle, RG trajectories (grey lines) should be flowing down straightly, but the scaling flows (red arrows) in Fig. \ref{FigRGt} (d) are not the case. We conjecture that when the inter-layer coupling is much larger than the intra-layer hopping, the in-plane electronic motions might be suppressed, which leads to some competition induced instability of in-plane electronic transport. This may have an non-ignorable impact on the in-plane transport, which was, however, only understood in a purely 2D regime in traditional RG analysis\cite{Pruisken1985}.

\begin{figure}[t]
\includegraphics*[width=0.47\textwidth]{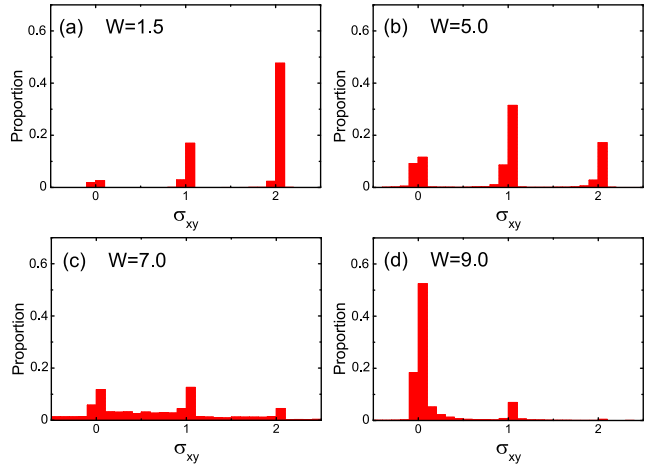}
\caption{ Similar to Fig. \ref{FigDistribute2} but corresponding to the case of Fig. \ref{FigRGt} (a) with $t=0.1$ and $A_1=0.3$, where the state with $\sigma_{xy}=1$ is a stable fixed point. }
\label{FigDistribute1}
\end{figure}

\subsection{C. Distributions of $\sigma_{xy}$ }

The above results were based on transport quantities averaged over disorder ensembles. To obtain more insights from another angle, now we investigate the statistical distributions of the Hall conductance at different disorder strength $W$. Fig. \ref{FigDistribute2} is the statistical histograms of $\sigma_{xy}$ at different disorder strength, corresponding to the scaling flows of Fig. \ref{FigRGA1} (a), with just one semicircle profile of scaling flows and with a weak $\sigma_{xy}=1$ state. On the other hand, Fig. \ref{FigDistribute1} corresponds to the case of Fig. \ref{FigRGt} (a), with two semicircles and with a visible $\sigma_{xy}=1$ plateau. When the disorder is weak, $W=1.5$ [panels (a) in both figures], both cases show isolated narrow peaks with quantized Hall conductances,
among which the $\sigma_{xy}=2$ peak is the most prominent. This suggests that the system is in a well defined insulating state with an extremely high probability of $\sigma_{xy}=2$. Such a distribution gives rise to a visible $\sigma_{xy}=2$ plateau after ensemble average, as we have seen above.

The most remarkable difference between Figs. \ref{FigDistribute2} and \ref{FigDistribute1} appears at the intermediate disorder strength $W=5$ [panels (b)]. In Fig. \ref{FigDistribute1} (b), similarly, an isolated and prominent narrow peak at $\sigma_{xy}=1$ results in a visible transport plateau of $\sigma_{xy}=1$ which is the stable fixed point in Fig. \ref{FigRGt} (a). On the other hand, the pattern is different in Fig. \ref{FigDistribute2} (b). Here although the $\sigma_{xy}=1$ peak is still very high, it is rooted in continuous and broad tails,
where $\sigma_{xy}$ is not quantized. This suggests that there exist considerable bulk states at this fermi level, making $\sigma_{xy}=1$ only a ``weak topological state'' which is hardly visible in a mesoscopic system, as stated above.

\subsection{D. Chern-Simons axion coupling}

\begin{figure}[htbp]
\includegraphics*[width=0.44\textwidth]{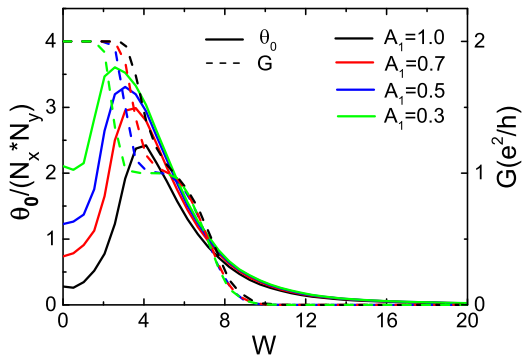}
\caption{(Color online) Solid curves: The area normalized finite size CSA coupling $\theta_0/(N_x N_y)$ as a function of disorder for different $A_1$. Each data point is an average over 4000 disorder configurations with size $N_x=N_y=20$. For comparison with the transport property, the corresponding two-terminal conductances are also plotted in dashed curves. The interlayer coupling $t=0.3$, and the other parameters
are same as Fig. \ref{FigDispersion} }
\label{FigthetaG}
\end{figure}

In this subsection, we explore how CSA coupling strength $\theta_{\mathrm{CS}}$ behaves during the process towards localization. Due to the size relation, Eq. (\ref{EqTheta02}), what makes sense for different sizes is the finite size CSA coupling (\ref{EqTheta0}) normalized by the sample area, $\theta_0/(N_x N_y)$. In Fig. \ref{FigthetaG},
solid curves are the $\theta_0/(N_x N_y)$ as a function of disorder strength $W$ for different $A_1$. For a comparison with the transport properties, we also plot the corresponding two-terminal conductance [identical to Fig. \ref{FigConduct} (a)] as dashed curves. It is interesting to notice that $\theta_0$ starts to increase sharply just before the collapse of the conductance plateau with $G=\sigma_{xy}=2$. Then it climbs to a peak after which the tendency of localization begins to dominate. However, the decay of $\theta_0$ is slower than that of $G$ after $W>10$. Another distinct feature is that there is also only one peak for all these cases, even for the case of $A_1=0.3$ where there are two visible transitions of the Hall conductance: $\sigma_{xy}=2\rightarrow 1$ and $1\rightarrow 0$. These facts suggest that during the process of localization, the variation of CSA coupling is closely related to the
disorder induced topological phase transition, but not a simple ``one to one'' correspondence with transport properties.

\begin{figure}[htbp]
\includegraphics*[width=0.38\textwidth]{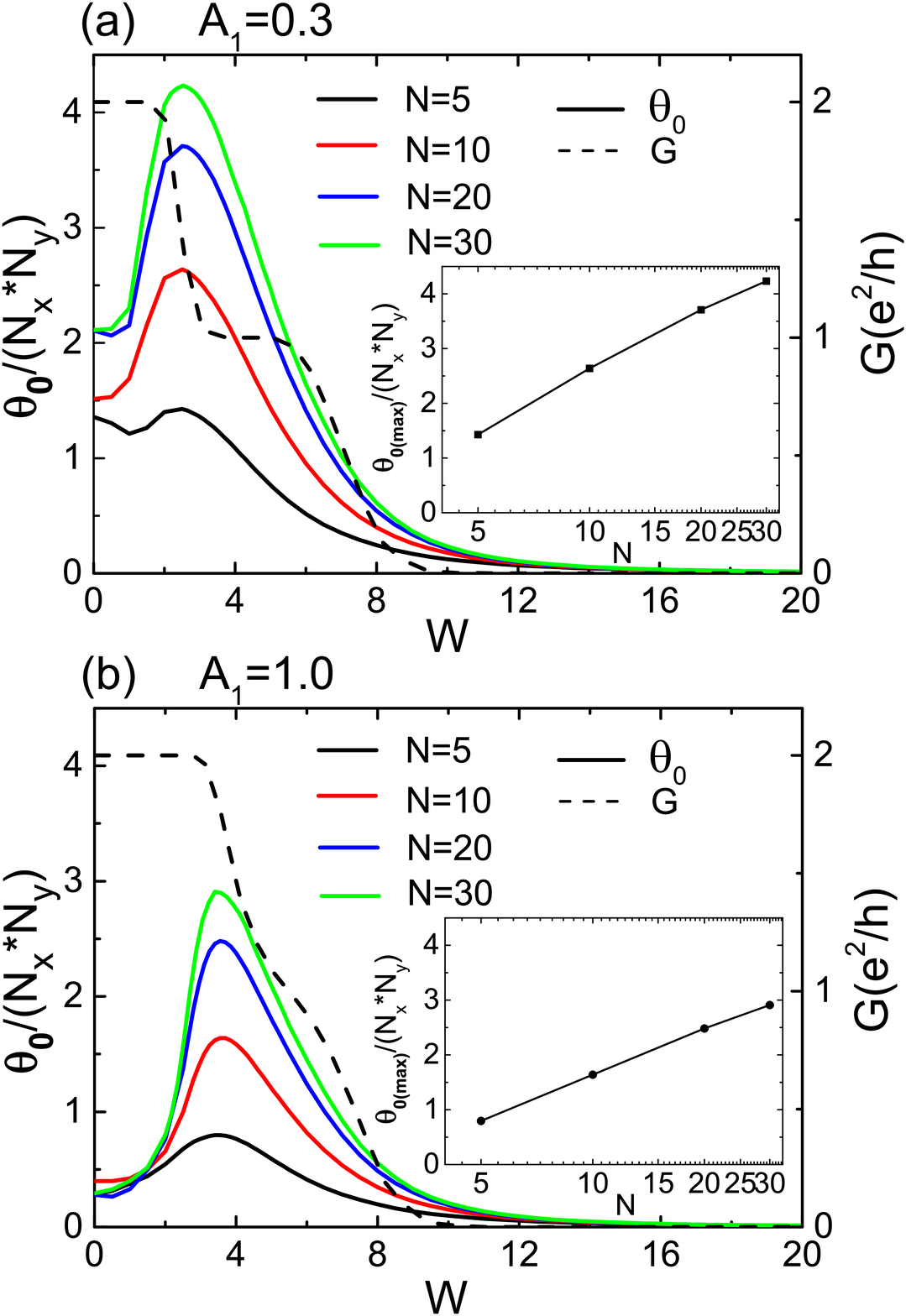}
\caption{(Color online) Similar to Fig. \ref{FigthetaG} but for different sample size $N\equiv N_x=N_y$, (a): $A_1=0.3$, (b): $A_1=1$. The insets are the corresponding peak value $\theta_{0(\mathrm{max})}$ as a function of sample size. }
\label{FigthetaN2}
\end{figure}

To obtain the information in the thermodynamic limit, we present the CSA coupling $\theta_0/(N_x N_y)$ for increasing sample sizes in Fig. \ref{FigthetaN2}, for $A_1=0.3$ and $A_1=1$ respectively. In both cases, the peak shapes of CSA become sharper with the increasing of the sample sizes.
The peak value $\theta_{0\mathrm{(max)}}/(N_x N_y)$ as a function of the sample size is plotted in the corresponding inset respectively. These peak values do not seem to converge until the largest size we can calculate, $N_x\times N_y=30\times30$. With the size axis in the logarithmic scale, one can see that its dependence on sample size $N$ is as fast as (or a little bit slower than) $\theta_{0\mathrm{(max)}}/(N^2)\sim \ln N$.
This scaling growth makes this CSA coupling experimental observation possible in realistic materials. Therefore,
besides intrinsic materials previously proposed\cite{Coh2011,XGWan2012,Mogi2017},
we find that multilayer QAH systems with disorder are also candidates for high $\theta$ materials.

The unexpected peak of $\theta_0$ prior to localization leads to a question about the underlying physics: how the electronic states are re-organized in the system during this process? To answer this, it is helpful to map $\theta_0$ to the real space. The zero-dimensional (finite sample) expression of $\theta_0$, Eq. (\ref{EqTheta0}), is believed to be a more fundamental definition of the CSA coupling than those $k$-space versions, and is free from the ambiguity of $2\pi$ and the subtleties of boundary conditions\cite{Olsen2017}. Therefore in the spirit of Refs. \cite{Resta2011,Resta2013,Resta2017}, it is reasonable to define the ``local marker'' of the CSA coupling for a real space site $\bm{r}$ as
\begin{equation}
\theta_0(\bm{r})=-8\pi^2\mathrm{Im}\mathrm{Tr}_{\bm{r}}[P\mathit{x}P\mathit{y}P\mathit{z}]\label{EqTheta0r},
\end{equation}
The only difference from Eq. (\ref{EqTheta0}) is that the trace $\mathrm{Tr}_{\bm{r}}$ is only over the freedom degrees within the site $\bm{r}$, which is two orbitals in our model (\ref{EqBHZ}). The sum of $\theta_0(\bm{r})$ over all $\bm{r}$ of the whole finite sample is just $\theta_0$.

\begin{figure}[t]
\includegraphics*[width=0.48\textwidth,bb=235 230 1640 2037]{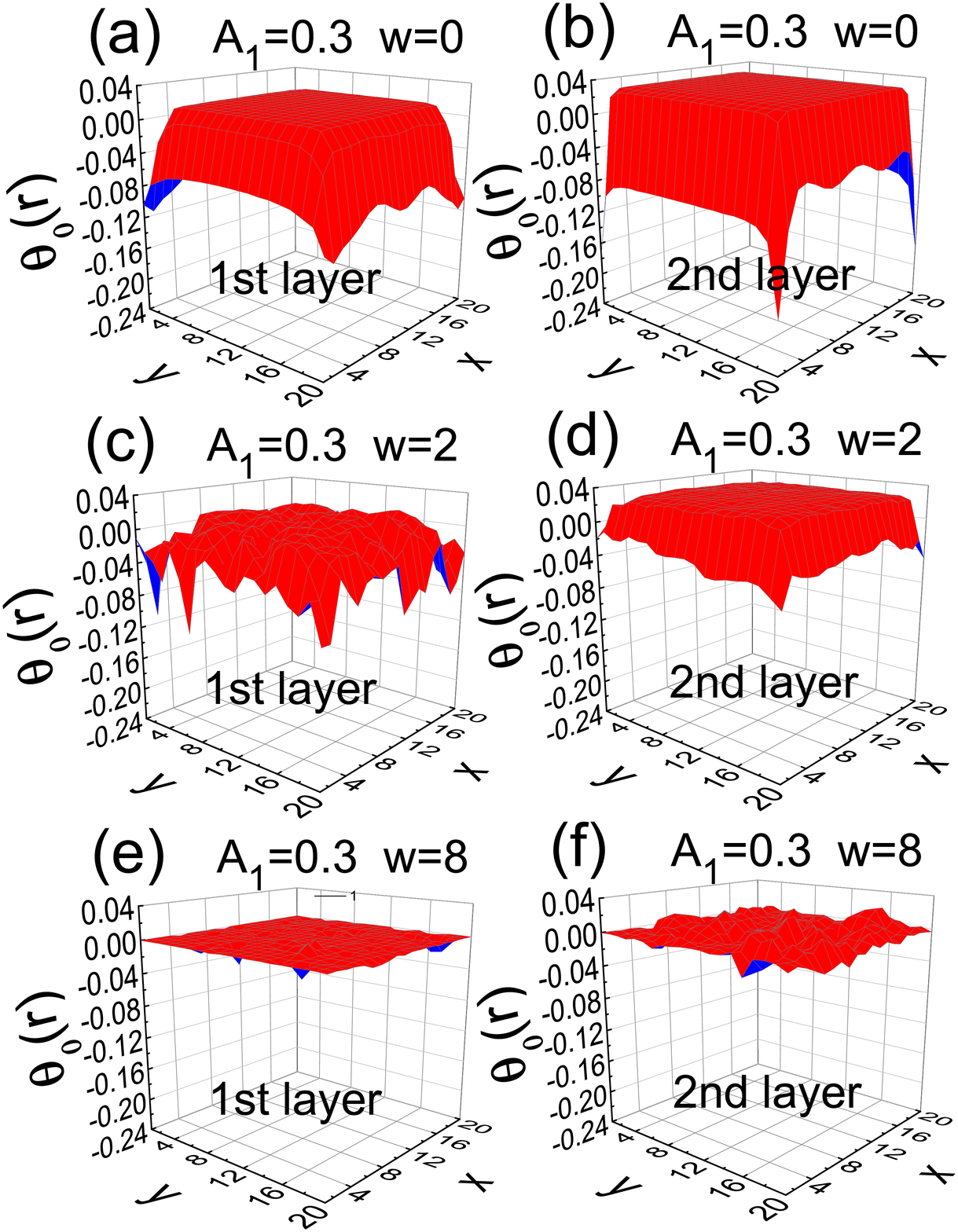}
\caption{(Color online) Real space distribution of $\theta_0(\bm{r})$ for a single typical disorder configuration with $A_1=0.3$, corresponding to the case of Fig. \ref{FigthetaN2} (a). The sample size is $20\times 20$. The left (right) column corresponds to $\theta_0(\bm{r})$ on the first (second) layer, and three rows correspond to three different disorder strengths $W$. }
\label{FigTheta0r03}
\end{figure}

\begin{figure}[t]
\includegraphics*[width=0.46\textwidth,bb=231 420 1720 2400]{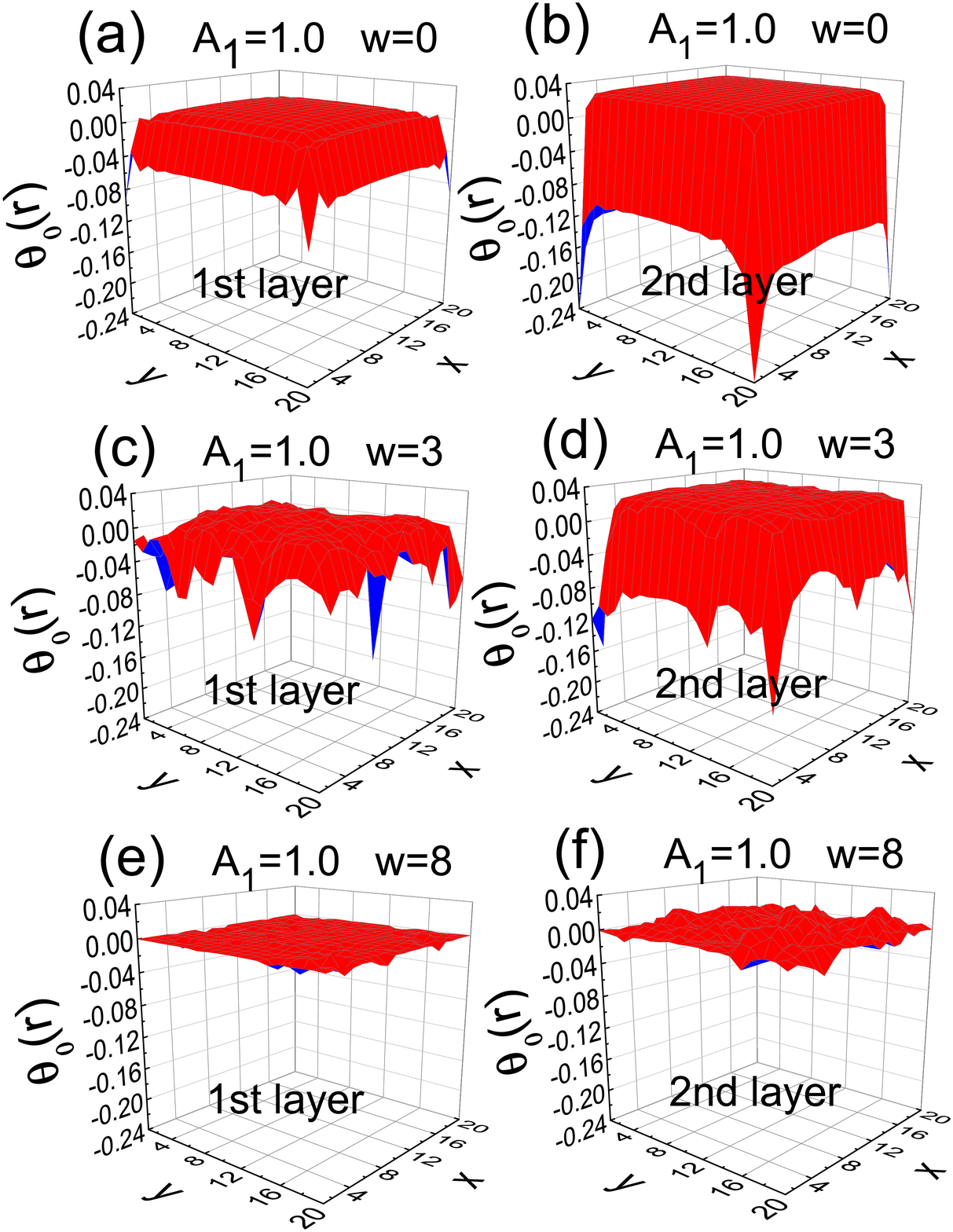}
\caption{(Color online) Similar to Fig. \ref{FigTheta0r03} but for $A_1=1$,
corresponding to the case of Fig. \ref{FigthetaN2} (b). }
\label{FigTheta0r10}
\end{figure}

To have a first impression, let us see the development of the real space distribution of the local marker $\theta_0(\bm{r})$ with the increasing of disorder strength $W$, for a single disorder configuration. Here, ``a single disorder configuration'' means that the onsite potentials (\ref{EqRandomOnsite}) of the sample have the form
\begin{equation}
U_{L;i}=W\cdot \varepsilon_{L;i},\label{EqSingleConfiguration}
\end{equation}
where $\{\varepsilon_{L;i}\}$ are a \emph{specific} realization of random numbers on $2 N_x N_y$ sites of the sample, which are uniformly distributed in $(-0.5,0.5)$, and $W$ is the \emph{single} variable to control the disorder strength. The results are shown in Figs \ref{FigTheta0r03} and \ref{FigTheta0r10}.

\begin{figure}[htbp]
\includegraphics*[width=0.45\textwidth]{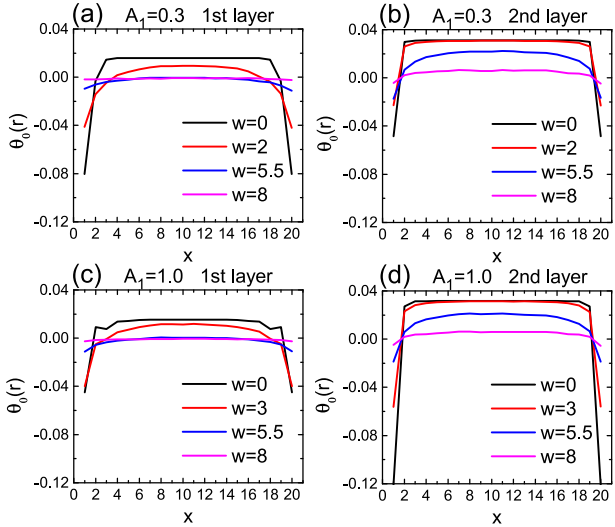}
\caption{(Color online) Disorder averaged $\theta_0(\bm{r})$ along the central cross section of $20\times20$ samples, for different disorder strength $W$. The upper (lower) row is for the case of $A_1=0.3$ ($A_1=0.3$), and the left (right) column corresponds to the first (second) layer. The average is over 5000 disorder configurations. }
\label{FigTheta0Average}
\end{figure}

Then afterwards, to smooth out disorder fluctuations and also for a more quantitatively and statistically reliable observation, similar to Ref. \cite{Resta2011,Resta2013}, we plot the disorder averaged $\theta_0(\bm{r})$ along the central cross section of the sample in Fig. \ref{FigTheta0Average}, with different colors for different disorder strengths.

In the absence of disorder, $W=0$, as shown in Figs. \ref{FigTheta0r03} and \ref{FigTheta0r10} [panels (a) and (b)], and in Fig. \ref{FigTheta0Average}, the local marker of the CSA coupling $\theta_{0}(\bm{r})$ on the edges is negative to that in the bulk, leading to a small positive value of total $\theta_0$ due to the compensation. Such negative contribution from the edges reflects the peculiar property of the edge states\cite{Resta2011,Resta2013}. With the increasing of disorder,
the typical magnitude of $\theta_0(\bm{r})$ on the edges is suppressed much faster than that in the bulk, as can be clearly seen in panels (c) and (d) of Figs. \ref{FigTheta0r03} and \ref{FigTheta0r10}. The statistically averaged data in Fig. \ref{FigTheta0Average} also tells the same story. For example in Fig. \ref{FigTheta0Average} (b) and (d) associated with the second layers  (which contribute $|\theta_0(\bm{r})|$ dominantly) in both cases, when $W$ is increased from $0$ (black curves) to $2$ (red curves), the local values of $\theta_0(\bm{r})$
in the bulk almost remain intact, but those on the edges are reduced by half. This reduction of compensation results in a rise of the total $\theta_0$ and a subsequent peak observed in Figs. \ref{FigthetaG} and \ref{FigthetaN2}.
Finally in the strong disorder limit, in panels (e) and (f), all types of orbital motions will be localized, and $\theta_0(\bm{r})$ tend to vanish uniformly.

\section{V. Summary}

In summary, the localization trajectories of the bilayer QAH systems with $C=2$ are investigated numerically. Distinct to
the monolayer counterpart in modular class $\Gamma_{\mathrm{R}}$\cite{ZG2015}, this bilayer system is in class $\Gamma_{\mathrm{T}}$, where there is a stable state with $C=1$ before the complete localization. When the bulk gaps associated with two layers are similar,
this intermediate state with $C=1$ is weak at finite size but still stable in the extremely large size limit. These pictures are confirmed by numerical simulations of two-parameter scaling flows on the conductance plane. However, some of the scaling flows deviate from the standard Pruisken pattern, especially in the case of strong inter-layer coupling. Microscopically, different performances of this intermediate state with $C=1$ correspond to different types of distribution of the Hall conductance, and the weakness of the intermediate state $C=1$ can be attributed to residual bulk states in the system. These numerical findings can be tested by state of the art experimental techniques\cite{RG_Exp2,Checkelsky2014,SGrauer2017} and raise new topics for renormalization group theories\cite{Pruisken1985,Olsen2017}. Finally, we find the CSA coupling $\theta$ of such systems can be significantly high in the transition window with medium disorder strength. This phenomenon is understood by a real space analysis of the corresponding local marker, by which we find that responses of the edge and the bulk to the disorder are different, and that the weak disorder leads to a suppression of edge-bulk compensation and gives rise to an increase of $\theta_0$. Therefore we propose a new way of finding candidate axion insulators.

\section{Acknowledgements}

This work was supported by National Natural Science
Foundation of China under Grant Nos. 11774336, 61427901 and 11374294. YYZ was also supported by the Starting Research Fund from Guangzhou University under Grant No. 69-18ZX10055.

\end{document}